\documentclass[conference]{IEEEtran}
\IEEEoverridecommandlockouts
\usepackage{graphicx}
\usepackage{subfigure}
\usepackage{amsmath,amsthm,amsfonts,amssymb}
\usepackage{enumerate}
\usepackage{cite}
\usepackage{bm}
\usepackage{bbm}
\usepackage{url}
\usepackage{array}
\usepackage{color,soul}
\usepackage{subcaption} 
\usepackage{subfigure} 
\usepackage{enumitem}
\usepackage{romannum}
\usepackage{amssymb}

\usepackage{booktabs}

\theoremstyle{plain}

\newtheorem{rem}{Remark}
\newtheorem{sty1}{Theorem}
\newtheorem{defi}[sty1]{Definition}

\allowdisplaybreaks[4]

\begin{document}
\title{Variable-Length Feedback Codes via Deep Learning}

\author{
Wenwei~Lai,
Yulin~Shao,
Yu~Ding,
Deniz~G\"und\"uz
\thanks{W. Lai, Y. Shao, and Y. Ding are with the State Key Laboratory of Internet of Things for Smart City and the Department of Electrical and Computer Engineering, University of Macau, Macau S.A.R. (e-mails: \{mc35291,ylshao,yc47469\}@um.edu.mo).}
\thanks{D. G\"und\"uz is with the Department of Electrical and Electronic Engineering, Imperial College London, London SW7 2AZ, U.K. (e-mail: d.gunduz@imperial.ac.uk).
}
}

\maketitle

\begin{abstract}
Variable-length feedback coding has the potential to significantly enhance communication reliability in finite block length scenarios by adapting coding strategies based on real-time receiver feedback. Designing such codes, however, is challenging. While deep learning (DL) has been employed to design sophisticated feedback codes, existing DL-aided feedback codes are predominantly fixed-length and suffer performance degradation in the high code rate regime, limiting their adaptability and efficiency. This paper introduces deep variable-length feedback (DeepVLF) code, a novel DL-aided variable-length feedback coding scheme. By segmenting messages into multiple bit groups and employing a threshold-based decoding mechanism for independent decoding of each bit group across successive communication rounds, DeepVLF outperforms existing DL-based feedback codes and establishes a new benchmark in feedback channel coding.
\end{abstract}

\section{Introduction}
Channel coding in the presence of feedback has been a pivotal research area in information and coding theory for decades \cite{shannon1956zero,schalkwijk1966coding1,kim2011error,polyanskiy2011feedback,ben2017interactive}. The classical feedback channel model, established by Shannon \cite{shannon1956zero}, involves data transmission from a transmitter to a receiver over a memoryless noisy channel, with the receiver providing real-time feedback to aid the forward channel coding. Although feedback does not increase the forward channel capacity, it significantly enhances communication reliability in the finite block length regime. A seminal example is the Schalkwijk-Kailath (SK) scheme \cite{schalkwijk1966coding1}, which achieves a double exponential decay of the block error rate with respect to the block length, underscoring the profound impact of feedback on error performance.

Designing feedback channel codes is considerably more intricate than designing forward channel codes due to the necessity for real-time adjustments based on the receiver's state \cite{polyanskiy2011feedback,ben2017interactive}. In forward channel coding, the coding strategy is fixed and depends only on the message, simplifying the code design. In contrast, feedback codes must dynamically adapt to the feedback, optimizing performance under varying channel conditions and receiver states. This complexity poses significant challenges in developing efficient and reliable feedback coding schemes.

In recent years, deep learning (DL) has emerged as a promising approach for designing sophisticated communication systems, including feedback channel codes \cite{kim2018deepcode,safavi2021deep,shao2023attentioncode,ozfatura2022all,ozfatura2023feedback,ankireddy2024lightcode,shao2024deep,vejling2023learning}. DL models can learn complex encoding and decoding structures directly from data, capturing intricate patterns and dependencies inherent in the communication and feedback processes. A pioneering effort in this domain is DeepCode \cite{kim2018deepcode}, where the authors proposed a  framework utilizing bit-by-bit passive feedback -- the receiver feeds back raw, noise-contaminated symbols without processing them. In DeepCode, recurrent neural networks (RNNs) at both the encoder and decoder create and exploit correlations among source bits and receiver feedback, improving decoding performance. 

AttentionCode \cite{shao2023attentioncode} introdu\-ced attention mechanism-based deep neural networks (DNNs) to replace RNNs, enabling the creation of longer-range correlations among source bits and feedback. In conjunction with a bit alignment mechanism, AttentionCode achieved significantly lower block error rates (BLERs) compared to DeepCode. The state-of-the-art was improved significantly with the introduction of Generalized Block Attention Feedback (GBAF) codes \cite{ozfatura2022all}, which partition message bits into small groups and perform group-level encoding and decoding using a seque\-nce-to-seq\-uence transformer architecture. This strategy efficiently reduces the input length to the DNN, thereby simplifying the learning process. Subsequent work \cite{ozfatura2023feedback} extended GBAF to active feedback scenarios, where the receiver processes the received signal before feeding it back to the transmitter. Additionally, LightCode \cite{ankireddy2024lightcode} was proposed as a lightweight coding architecture that achieves performance comparable to GBAF but with significantly fewer learning parameters.

Despite these advancements, existing DL-aided feedback codes are predominantly fixed-length, which limits their adaptability and may prevent them from fully exploiting the benefits of feedback. Moreover, these codes perform well in the low code rate regime -- typically lower than $1/3$ -- but their performance deteriorates significantly as the code rate increases, where fewer channel uses are available for transmission.

{\it Contributions:} To address these challenges, this paper introduces a novel DL-aided variable-length feedback code, termed deep variable-length feedback (DeepVLF) code. DeepVLF leverages multi-round feedback to achieve adaptive code rates with bit-group-level granularity, allowing the transmitter to adjust the amount of information transmitted based on the feedback. By dividing the message into bit groups and employing a threshold-based decoding strategy, each bit group can be decoded independently across different communication rounds. This approach reduces the total number of channel uses required to achieve the same performance as fixed-rate codes, particularly in high rate regimes.
To jointly optimize the encoder and decoder, we develop a two-phase training strategy with a custom-designed loss function tailored to minimize the error rate of bit groups while adhering to power constraints. Numerical experiments demonstrate that DeepVLF sets a new state-of-the-art in feedback channel coding, especially in the high code rate regime.

\section{System Model}
\begin{figure}[t]
    \centering
    \includegraphics[width=1\linewidth]{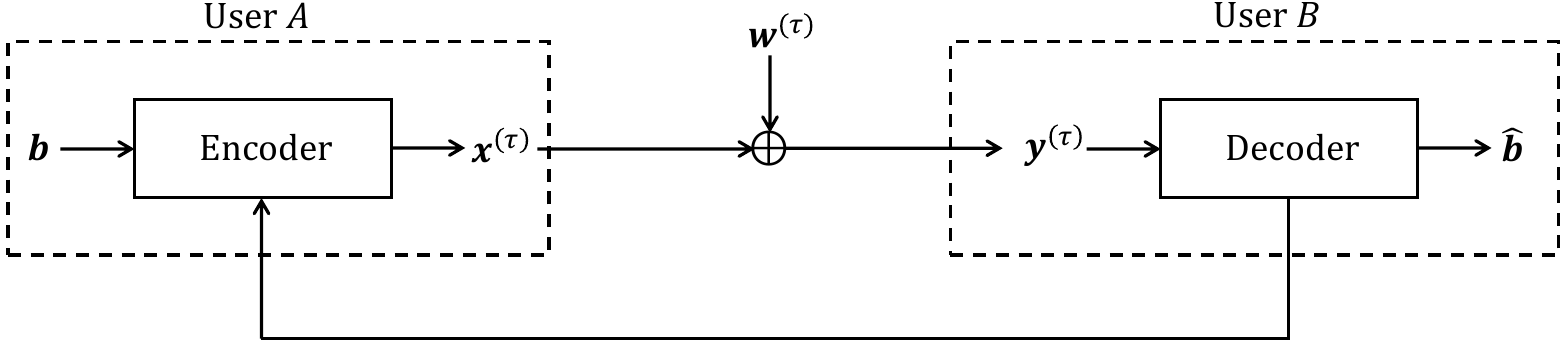}
    \caption{System model of feedback channel coding.}
    \label{fig:system}
\end{figure}

We consider communications from user A to user B in the presence of feedback, as illustrated in Fig. \ref{fig:system}. Both forward and feedback channels are modeled as additive white Gaussian noise (AWGN) channels. In practice, user B often represents a base station (BS) that possesses greater transmission power and employs multiple antennas. This enhanced capability allows the BS to achieve highly reliable communication over the feedback link. As a result, we consider the feedback channel to be noiseless -- a common setup in the literature to fully exploit the capabilities of feedback.

Let $\bm{b}\in\{0,1\}^{K\times 1}$ denote the $K$ information bits to be communicated from user A to B. With feedback from user B, the communications take place in an interactive fashion, and can be organized into multiple rounds, each consisting of a forward transmission from A to B, followed by a feedback response from B to A. This cyclical interaction allows for continual adjustments to the coding strategy of user A based on the feedback, gradually increasing the likelihood of successful transmission.

In the forward transmission phase of communication round $\tau$, user A constructs a packet of parity symbols $\bm{x}^{(\tau)}$ utilizing both the information bits $\bm{b}$ and the feedback received from previous rounds. This packet is then transmitted over the forward AWGN channel to user B, and the received signal can be written as
\begin{equation}
    \bm{y}^{(\tau)} = \bm{x}^{(\tau)} + \bm{w}^{(\tau)},
\end{equation}
where $\bm{w}^{(\tau)}$ is an AWGN vector with independent and identical distributed (i.i.d) component adheres to a Gaussian distribution $\mathcal{N}(0,\sigma^2)$. Throughout the paper, we use the superscript $\tau$ to denote the communication round.

Over the feedback channel, user B can transmit the received symbols $\bm{y}^{(\tau)}$ back to user A. This allows user A to analyze the noise realizations that have affected the transmitted symbols $\bm{x}^{(\tau)}$ in the forward channel. With this knowledge, user A can then refine the encoding strategy in future rounds to minimize the impact of noise. Consequently, the fundamental challenge in feedback channel coding lies in the strategic design of both the encoder and decoder to utilize this feedback effectively, enhancing the overall reliability and efficiency of the communication process.

\begin{rem}
The DeepVLF code proposed in this paper belongs to a class of DL-aided feedback channel codes, which are often robust to noisy feedback. While the primary focus of this work is on a noiseless feedback setting, our numerical experiments demonstrate that DeepVLF maintains superior performance under noisy feedback conditions as well. This confirms that DeepVLF consistently outperforms existing codes, even in the presence of feedback noise. 
\end{rem}

\section{Methodology}
This section outlines the architecture of the DeepVLF code. We discuss the design and functionality of both the encoder and decode, highlighting the innovative aspects that differentiate our approach from previous designs.

\subsection{Key elements}
At the core of the DeepVLF architecture are several fundamental elements that guide its design.

\begin{defi}[bit groups]
The encoding and decoding of DeepVLF are organized around groups of bits instead of individual bits. Given a vector of message bits $\bm{b}$, DeepVLF first divides $\bm{b}$ into $Q$ equal-size groups, with each group containing $m\triangleq K/Q$ bits:
\begin{equation}
    \bm{b} = [\bm{b}^\top_1,\bm{b}^\top_2,\bm{b}^\top_3,\cdots,\bm{b}^\top_Q]^\top,
\end{equation}
where each group $\bm{b}_q\in\mathcal{A}\triangleq\{0,1\}^{m\times 1}$.
\end{defi}

The total number of possible patterns per group is $|\mathcal{A}|=2^m$. This grouping approach is analogous to quadrature amplitude modulation (QAM), where symbols are formed by grouping multiple bits. By reducing the effective length of the input sequence for the DNN, this method efficiently reduces the computational and memory demands on the network \cite{ozfatura2022all}, facilitating more effective learning and processing.

\begin{defi}[belief matrix]
In the DeepVLF framework, user B formulates a belief matrix $\bm{P}$ to represent each transmitted bit group:
\begin{equation}\label{eq:belief}
    \bm{P} = [\bm{p}_1, \bm{p}_2, \bm{p}_3, \cdots, \bm{p}_Q],
\end{equation}
where column $\bm{p}_q$ corresponds to bit group $\bm{b}_q$. Specifically, $\bm{p}_q$ is a $2^m$-dimensional probability distribution, with the $j$-th element indicating the likelihood that user B interprets the transmitted bit group $\bm{b}_q$ as the $j$-th element of set $\mathcal{A}$.
\end{defi}

\begin{defi}[threshold decoding]\label{defi:thres_dec}
In DeepVLF, user B utilizes a threshold decoding strategy to decode each bit group individually. Specifically, a bit group $\bm{b}_q$ is considered successfully decoded by user B if any element of the belief vector $\bm{p}_q$ surpasses a predefined threshold $\gamma$, i.e.,
\begin{equation}\label{eq:thres}
\|\bm{p}_i\|_{\infty}\geq \gamma,
\end{equation}
where $\|\cdot\|_{\infty}$ denotes the infinity norm, which evaluates the maximum absolute value among the elements of a vector.

Contrary to existing code designs, where decoding is performed only after the complete sequence of transmissions are received by the decoder, DeepVLF employs threshold decoding at the end of each communication round, and iteratively updates its belief with the help of the feedback link.
\end{defi}

\begin{defi}[feedback information]
In DeepVLF, during each communication round, user B provides user A with two types of feedback information:
\begin{itemize}
    \item The packet of received symbols $\bm{y}^{(\tau)}$. 
    \item The indices of the bit groups that have been successfully decoded within the last communication round.
\end{itemize}
\end{defi}

\subsection{Encoder}
The encoding process of DeepVLF is iterative, unfolding across several rounds. Each round leverages feedback to refine the generation of parity symbols, thereby enhancing the robustness and efficiency of the transmission. This section details the mechanics of how feedback is incorporated within each round and the subsequent generation of parity symbols.

\begin{figure}[t]
    \centering
    \includegraphics[width=0.65\linewidth]{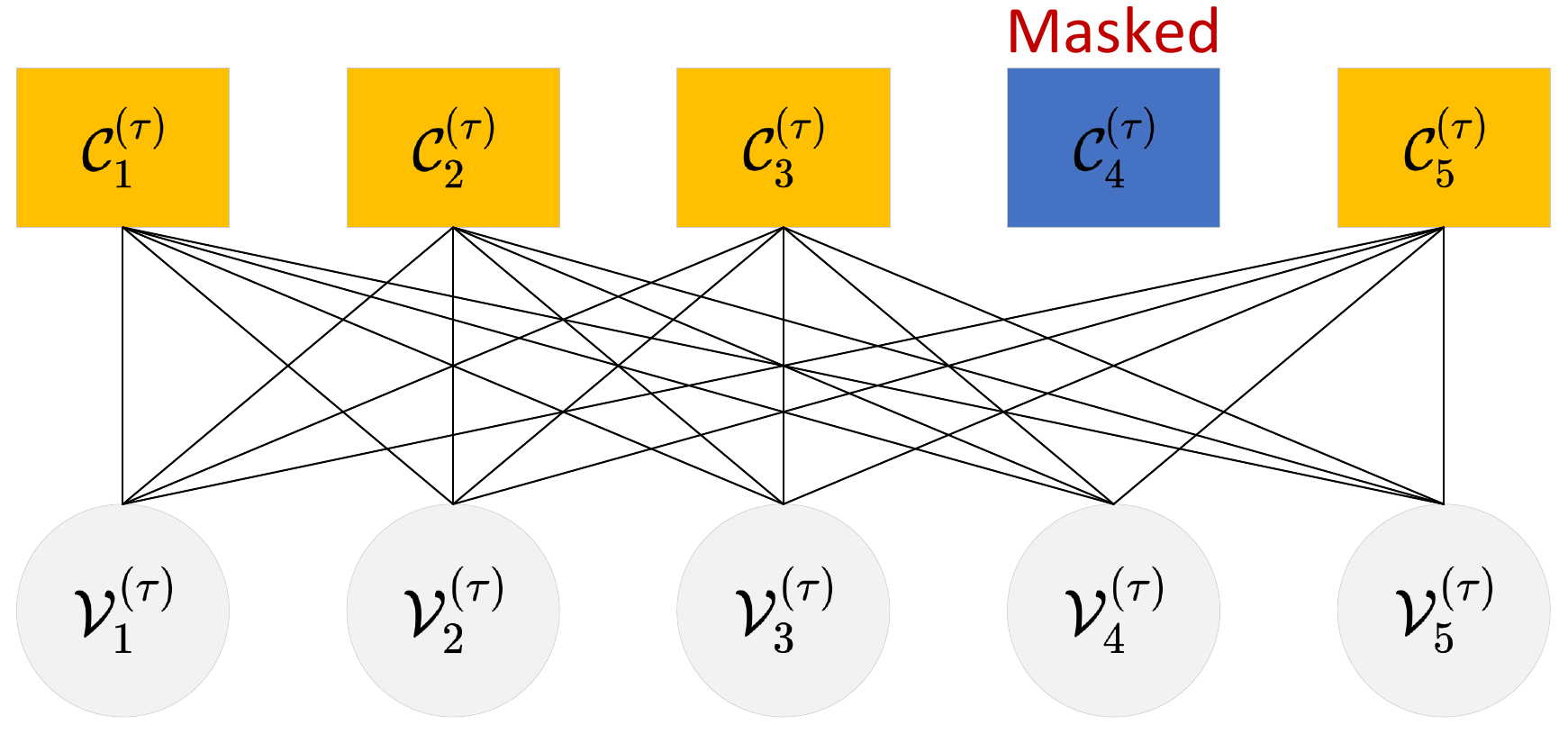}
    \caption{The generation of parity symbols in each communication round can be represented by a bipartite graph.}
    \label{fig:graph}
\end{figure}

During each communication round $\tau$, the generation of parity symbols $\bm{x}^{(\tau)}$ can be represented by a bipartite graph with variable nodes $\mathcal{V}^{(\tau)}_i$, $i=1,2,...,Q$, check nodes $\mathcal{C}^{(\tau)}_i$, $i=1,2,...,Q$, and a set of edges $\mathcal{E}^{(\tau)}$, as shown in Fig.~\ref{fig:graph}.
\begin{itemize}
    \item Each $i$-th variable node and check node is uniquely associated with the $i$-th bit group, designed specifically to cater to the encoding and decoding needs of that group. This ensures targeted processing of each bit group throughout the communication rounds.
    \item Each check node contains a parity symbol intended for transmission to user B. As stated in Definition \ref{defi:thres_dec}, DeepVLF actively decodes the bit groups in all interactions. If bit group $i$ is decoded successfully during communication round $\tau$, the parity symbol in $\mathcal{C}^{(\tau^\prime)}_j$, $\tau^\prime>\tau$, will no longer be transmitted to user B.
    \item The variable nodes contain the information to generate the check nodes. Initially, in the first round, we have $\mathcal{V}^{(1)}_i=\{\bm{b}_i\}$. With each subsequent communication round, $\mathcal{V}^{(\tau)}_i$ is updated to incorporate additional feedback from user B, enriching the information used for future encoding decisions and enhancing the system's adaptability to channel noise realizations.
\end{itemize}

The check nodes and variable nodes are interconnected through a set of edges. Let $e_{ij}$ represent a possible edge linking the $i$-th variable node with the $j$-th check node. During communication round $\tau$, the set of active edges is denoted by $\mathcal{E}^{(\tau)}$. The configuration of $\mathcal{E}^{(\tau)}$ is dynamically adjusted based on the decoding status of bit groups as follows:
\begin{itemize}
    \item Initially, when no bit group has been decoded, all possible connections are active, thus $\mathcal{E}^{(\tau)}=\{e_{ij},\forall i, j\}$.
    \item If a bit group $j^\prime$ was successfully decoded in the previous round $\tau-1$, the edges connecting to the corresponding check node $j^\prime$ are deactivated in the current round. Therefore, $\mathcal{E}^{(\tau)}=\mathcal{E}^{(\tau-1)}\backslash\{e_{ij^\prime},\forall i\}$.
    \item Each edge $e_{ij}\in \mathcal{E}^{(\tau)}$ is associated with a weight $\rho_{ij}$. For each $\mathcal{C}_j$, we have $\sum_{i=1}^{Q} \rho_{ij} = 1$.
\end{itemize}

Given the bipartite graph,  the relationship between the variable and check nodes can be written as
\begin{equation}\label{eq:enc}
    \mathcal{C}_j = f_c\Bigg(\sum_{i=1}^{Q} \rho_{ij} f_v(\mathcal{V}_i)\Bigg),
\end{equation}
where the encoding functions $f_c$ and $f_v$ are realized by DNNs. The coefficients $\rho_{ij}$ are learnable parameters, which are optimized to effectively determine the influence of each variable node on the corresponding check node. This architecture of DNNs and their role in mapping variable nodes to check nodes is detailed further in Section \ref{sec:IV}.

The updating process for variable nodes after each communication round is crucial for the overall system's adaptability and efficiency. This update is governed by an auto-regressive mechanism as described below:
\begin{equation}\label{eq:vninfo}
    \mathcal{V}^{(\tau)}_i =
    \begin{cases}
     \{\bm{b}_i\},& \text{if $\tau=1$;}\\
     \mathcal{V}^{(\tau-1)}_i \cup \mathcal{C}^{(\tau-1)}_i \cup \{\bm{y}^{(\tau-1)}_i\},& \text{if $\|\bm{p}^{(\tau-1)}_i\|_{\infty}<\gamma$;}\\
     \mathcal{V}^{(\tau-1)}_i,& \text{otherwise.}
    \end{cases}
\end{equation}

In the initial round ($\tau=1$), each variable node is simply initialized to the corresponding bit group. In subsequent rounds, the update depends on the decoding success: if the infinity norm of the probability vector $\bm{p}^{(\tau-1)}_i$ is below the threshold $\gamma$, indicating unsuccessful decoding, $\mathcal{V}^{(\tau)}_i$ is expanded to include the parity information from the last round and the received symbols. If the threshold is met or exceeded, further updates are unnecessary, and $\mathcal{V}^{(\tau)}_i$ retains its prior state from $\mathcal{V}^{(\tau-1)}_i$. Nevertheless, this ``frozen'' node will continue to contribute to the generation of other parity symbols for the remaining rounds.

\subsection{Decoder}
At user B, the decoder engages in decoding operations during each communication round upon receiving $\bm{y}^{\tau}$. The design principles of the decoder mirror those of the encoder. In the following, we succinctly outline the decoder's functionality.

The core functionality of the decoder within DeepVLF is estimating the belief matrix $\bm{P}$. This process can also be modeled by a bipartite graph representation, consisting of variable nodes $\widetilde{\mathcal{V}}^{(\tau)}_i$, and check nodes $\widetilde{\mathcal{C}}^{(\tau)}_j$, akin to the encoder's structure in \eqref{eq:enc}. We have
\begin{equation}\label{eq:dec}
    \widetilde{\mathcal{C}}_j = \widetilde{f}_c\Bigg(\sum_{i=1}^{Q} \widetilde{\rho}_{ij} \widetilde{f}_v(\widetilde{\mathcal{V}}_i)\Bigg).
\end{equation}
In \eqref{eq:dec}, the variable node $\widetilde{\mathcal{V}}_i$ stores the received parity symbols from the previous communication rounds and the current belief about the $i$-th bit group $\bm{p}^{(\tau-1)}_i$. The check node $\widetilde{\mathcal{C}}_j$ estimates the updated belief $\bm{p}^{(\tau)}_j$ for the $j$-th bit group. The functions $\widetilde{f}_c$ and $\widetilde{f}_v$ are realized by DNNs, and $\widetilde{\rho}_{ij}$ are learnable parameters, mirroring the learnable weights $\rho_{ij}$ used in the encoder.

If $\|\bm{p}^{(\tau)}_i\|_{\infty}\geq \gamma$, the $i$-th bit group is deemed successfully decoded. Consequently, the corresponding $\widetilde{\mathcal{V}}^{(\tau)}_i$ and $\widetilde{\mathcal{C}}^{(\tau)}_j$ are frozen and will not undergo further updates in future communication rounds, thereby preserving the state achieved at the point of successful decoding.

\begin{defi}[code rate]
DeepVLF decodes bit groups individually across different communication rounds, thus introducing variability in the decoding timeline for each bit group. We define $\tau^*_q$ as the communication round during which the $q$-th bit group is decoded:
\begin{equation}
    \tau^*_q \triangleq \big\{\tau: \|\bm{p}^{(\tau-1)}_q\|_{\infty} < \gamma, \|\bm{p}^{(\tau)}_q\|_{\infty} \geq \gamma \big\}.
\end{equation}
The overall code rate $R$ of DeepVLF can be written as
\begin{equation}
    R = \frac{K}{\sum_{q=1}^Q \tau^*_q}.
\end{equation}  
\end{defi}

Given the DeepVLF framework described above, our primary objective is to maximize the code rate $R$ while ensuring that the block error rate (BLER) $\xi\triangleq\Pr(\bm{b} \neq \widehat{\bm{b}})$, where $\widehat{\bm{b}}$ denotes the decoded bit sequence, remains below a specified target $\xi^*$. This optimization is performed under an average power constraint. Formally, we have
\begin{equation*}
\begin{aligned}
& \max_{f_c,f_v,\widetilde{f}_c,\widetilde{f}_v, \{\rho_{ij},\widetilde{\rho}_{ij}\}}
& & R \\
& \hspace{1cm} {s. t.}
& & \xi \leq \xi^*,\\
& 
& & \mathbb{E} \left[ \frac{1}{\sum_{q=1}^Q \tau^*_q} \sum_{q=1}^Q\sum_{\tau=1}^{\tau^*_q}  \big(x_q^{(\tau)}\big)^2 \right] \leq 1.
\end{aligned}
\end{equation*}


\section{Realization of DeepVLF}\label{sec:IV}
This section presents our realization of the DeepVLF methodology using a transformer-based DNN architecture \cite{vaswani2017attention}.
The source codes and well-trained DNN models are available online.\footnote{Source code: \url{https://github.com/lynshao/DeepVLF}.}

\subsection{Parity symbol generation}
The encoding process at user A is guided by the principles outlined in \eqref{eq:enc}, wherein the variable nodes $\mathcal{V}^{(\tau)}_i$ are updated in each communication round $\tau$ based on the feedback received from user B following \eqref{eq:vninfo}. The knowledge vectors contained in the variable nodes are first projected onto a high-dimensional latent space through the feature extractor function $f_v$, implemented via a variable-depth feature extractor (VDFE). Subsequently, these latent vectors are aggregated using coefficients ${\rho_{ij}}$ generated by self-attention layers. The aggregated latent representations are then mapped to the check nodes $\mathcal{C}^{(\tau)}_j$ via a header function $f_c$, resulting in the generation of parity symbols for transmission.

\textbf{VDFE ($f_v$)}:
In DeepVLF, the feature extractor $f_v$ plays a crucial role in projecting the knowledge vectors of the variable nodes into a latent space suitable for processing by self-attention. We implement $f_v$ using fully-connected layers with ReLU activation functions. Notably, the depth of the fully-connected layers varies depending on the communication round $\tau$ through the introduction of a VDFE, where the number of fully-connected layers in $f_v$ is adjusted based on whether the current communication round $\tau$ is before or after a predefined threshold $\tau_{\text{VD}}$, while a shallower feature extractor suffices in the earlier rounds of communication ($\tau \leq \tau_{\text{VD}}$). In contrast, during the later rounds ($\tau > \tau_{\text{VD}}$), a deeper feature extractor is employed to model the increasing complexity and nonlinearity of the encoding function.

\begin{rem}
Our VDFE design is inspired by observations in existing DL-aided feedback codes, where the nonlinearity in the functional relationship between the input and output of the encoder increases as the communication rounds progress. An ablation study is provided in Section \ref{sec:sim} to verify the effectiveness of VDFE.
\end{rem}

\textbf{Self-attention coefficients $(\rho_{ij})$}:
The aggregation coefficients $\{\rho_{ij}\}$ determine the impact of each variable node on the check nodes during the encoding process. In DeepVLF, we employ a self-attention mechanism to dynamically generate these coefficients, leveraging the ability of attention mechanisms to capture dependencies and interactions between elements in a sequence. The effectiveness of using self-attention in feedback coding has been demonstrated in our prior works \cite{shao2023attentioncode,ozfatura2022all,ozfatura2023feedback}.

In this approach, each coefficient $\rho_{ij}$ is computed based on the inner product of the latent representations of the variable nodes:
\begin{equation} \rho_{ij} = \text{softmax}\left( f_v(\mathcal{V}_i^{(\tau)})^\top f_v(\mathcal{V}_j^{(\tau)}) \right), \end{equation}
where softmax ensures that the coefficients are non-negative and sum to one for each check node.

To realize the threshold decoding mechanism and account for the decoding status of bit groups, we utilize masking within the self-attention layers. Specifically, we define a mask $\mathcal{M}^{(\tau)} \in \{0,1\}^Q$ for the $Q$ bit groups, representing the decoding status at communication round $\tau$:
\begin{equation}
\mathcal{M}^{(\tau)}_j =
\begin{cases}
1, & \text{if } \|\bm{p}^{(\tau)}_j\|_{\infty}<\gamma\\
0, & \text{otherwise}
\end{cases}
\end{equation}

Thanks to the feedback channel, this mask is known to both the encoder and decoder. In the self-attention computation, we mask the check nodes whose associated bit group has been successfully decoded, as illustrated in Fig.~\ref{fig:graph}.

\textbf{Feedforward Network ($f_c$)}:
After the latent representations are aggregated using the self-attention coefficients, they are mapped to the check nodes through the header function $f_c$. We design $f_c$ as a two-layer fully connected network with Gaussian error linear unit (GeLU) activation functions \cite{hendrycks2016gaussian}. 

At the encoder side, the header function $f_c$ maps each aggregated latent vector to a single parity symbol $x^{(\tau)}_i$, which is then transmitted over the forward channel to user B. At the decoder side, the corresponding header function $\widetilde{f}_c$ generates a belief vector $\bm{p}^{(\tau)}_j$ for the $j$-th bit group by adding an additional linear layer followed by a softmax activation function. This setup allows the decoder to produce a probability distribution over the possible bit group patterns, facilitating effective threshold decoding as per Definition \ref{defi:thres_dec}.


\begin{figure*}[t]
    \centering
    \includegraphics[width=1\linewidth]{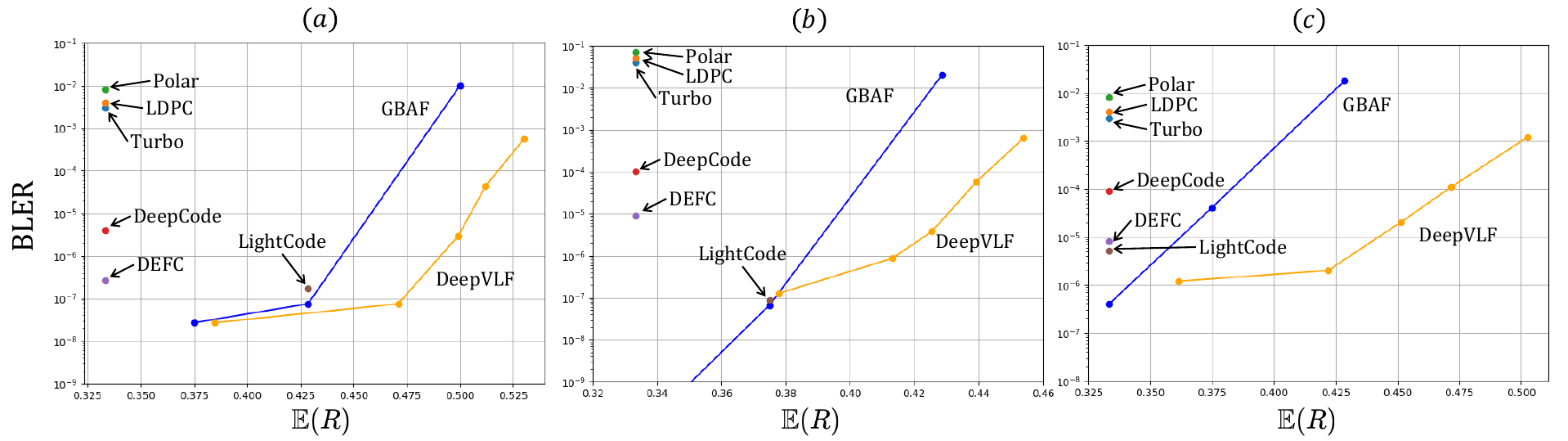}
    \caption{The BLER versus average code rate performances of DeepVLF: (a) forward SNR is 1 dB, noiseless feedback; (b) forward SNR is 0 dB, noiseless feedback; (c) forward SNR is 1 dB, feedback SNR is 20 dB.}
\label{fig:bler}
\end{figure*}

\subsection{Decoding and loss function design}
At user B, the decoder utilizes a similar transformer-based architecture to process the received signals and update the belief matrix $\bm{P}$. The decoder's variable nodes $\widetilde{\mathcal{V}}^{(\tau)}_i$ store the received parity symbols from previous communication rounds and the current belief about each bit group. The check nodes $\widetilde{\mathcal{C}}^{(\tau)}_j$ are responsible for generating the updated belief vectors $\bm{p}^{(\tau)}_j$, as described in \eqref{eq:dec}.

The decoder employs the same VDFE and self-attention mechanism, ensuring consistency in processing and facilitating effective learning of the encoding-decoding relationship. The mask vector $\mathcal{M}^{(\tau)}$ is also utilized at the decoder side to mask decoded bit groups.

Upon receiving the parity symbols $\bm{y}^{(\tau)}$, the decoder updates the belief vectors and applies the threshold decoding criterion. If $|\bm{p}^{(\tau)}_j|_{\infty} \geq \gamma$, the $j$-th bit group is declared successfully decoded, and its corresponding variable and check nodes are frozen in subsequent rounds. This iterative process continues until all bit groups are decoded.

Given that DeepVLF attempts to decode the bit groups after each communication round, the training process must accommodate the dynamic nature of the decoding accuracy, which improves over successive rounds. In the following, we elaborate on the loss function designed for DeepVLF and discuss the strategies employed to enhance its training efficacy.

DeepVLF is trained in a supervised manner, aiming to minimize the discrepancy between the transmitted bit groups $\bm{b}_i$ and the estimated probabilities $\bm{p}_i^{(\tau)}$ at each round $\tau$. A straightforward loss function design is to sum the cross-entropy losses for every bit group across all communication rounds:
\begin{equation}\label{eq:loss1}
    \mathcal{L} = - \sum_{q=1}^{Q}\sum_{\tau=1}^{\tau^*_q}  \sum_{j=1}^{2^m} \mathbb{I}_{(\bm{b}_q=\mathcal{A}_j)}\cdot\log(p_{q,j}^{(\tau)}),
\end{equation}
where $\mathbb{I}$ is the indicator function, and $\mathcal{A}_j$ denotes the $j$-th element in the set $\mathcal{A}$.
This loss function aims to minimize the cross-entropy between the true bit groups and the estimated probabilities at each round, encouraging the model to improve its decoding accuracy progressively.

However, due to the interactive nature of DeepVLF, the BLER naturally decreases as the communication rounds progress. In the initial rounds, the decoder's estimates are less accurate, resulting in higher cross-entropy losses. Consequently, when averaging the cross-entropy equally across all rounds, the loss function in \eqref{eq:loss1} becomes dominated by the high losses from the early rounds. This dominance can hinder the training process, as the model may focus disproportionately on minimizing the loss in the early rounds, potentially at the expense of performance in later rounds where significant improvements occur.

To mitigate this problem and enhance the performance of DeepVLF, we introduce two advancements in our training.

First, we set a threshold $\tau^+$, and initiate decoding attempts only after communication round $\tau^+$. Communication rounds before $\tau^+$ are treated as undecodable, acknowledging that the decoder's estimates during these early rounds are not sufficiently reliable. By excluding the high-loss contributions from the initial rounds, we prevent them from dominating the training process. The value of $\tau^+$ is determined based on the Shannon limit \cite{shannon1948mathematical}, providing a rough guideline for the minimum number of transmissions required for reliable communication given the channel conditions:
\begin{equation}\label{eq:tau+}
    \tau^+ \triangleq \max\bigg\{\mu, \Big\lfloor\frac{2m}{\log_2{\left(1 + \eta\right)})}\Big\rfloor\bigg\},
\end{equation}
where $\eta = \frac{1}{\sigma^2}$ is the signal-to-noise ratio (SNR) of the forward channel and $\mu$ is served as  a baseline threshold for $\tau^+$, adjusted based on $\gamma$.

Second, we incorporate an exponential weighting factor into the loss function to balance the contributions of losses from different rounds. By assigning higher weights to losses from later rounds -- where the decoder's estimates are more accurate -- we encourage the model to focus on improving performance in these critical stages. The final loss function for training DeepVLF is thus formulated as:
\begin{equation}\label{eq:loss2}
    \mathcal{L} = - \sum_{q=1}^{Q}\sum_{\tau=\tau^+}^{\tau^*_q} \vartheta^{\tau-\epsilon} \sum_{j=1}^{2^m} \mathbb{I}_{(\bm{b}_q=\mathcal{A}_j)}\cdot\log(p_{q,j}^{(\tau)}),
\end{equation}
where $\vartheta$ is the base of the exponential weighting factor while $\epsilon$ denotes a fixed offset parameter. By carefully selecting $\vartheta$ and $\epsilon$, we can effectively balance the influence of each communication round on the total loss, ensuring that later rounds -- more indicative of the model's ultimate performance -- have a greater impact on the training process. An ablation study comparing different loss functions is given in Section~\ref{sec:sim}.

\section{Numerical Experiments}\label{sec:sim}

This section evaluates the performance of DeepVLF benchmarked against existing feedback coding schemes.

\subsection{Training DeepVLF}
In our training procedure, we employ the AdamW optimizer, a refined version of the Adam optimizer that decouples weight decay from the gradient-based updates \cite{loshchilov2017decoupled}, and a learning rate scheduling strategy where the learning rate progressively decreases as the training steps advance. 

The average code rate of DeepVLF varies with the decoding threshold $\gamma$, which determines when bit groups are considered successfully decoded based on the belief matrix. To optimize DeepVLF's performance across different code rates and BLER requirements, we structure the training process into two phases:
\begin{itemize}
    \item \textbf{Pretraining:} The DeepVLF model undergoes pretraining with varying values of the decoding threshold $\gamma$. By exposing the model to a range of decoding conditions, it learns to capture diverse feature representations and adapts to different requirements of belief for decoding. This phase enhances the model's generalization capabilities and prepares it for various operational scenarios.
    \item \textbf{Fine-tuning:} Following pretraining, we fine-tune the model with a target value of $\gamma$ that aligns with the desired balance between the code rate and BLER. This phase focuses on optimizing the model's parameters to control DeepVLF's average code rate and to improve its BLER performance.
\end{itemize}

\begin{table}[t]
\caption{Hyperparameter settings for DeepVLF.}
\centering
\setlength{\tabcolsep}{3mm} 
\begin{tabular}{ccc}
\toprule
\textbf{Parameters} & \textbf{Descriptions} &\textbf{Value} \\ 
\midrule
$K$ &Bitstream length& $51$ \\ 
$Q$ &Number of bit groups& $17$ \\ 
$m$ &Size of a bit group& $3$ \\
$B$ &Batchsize& $8192$ \\ 
lr &Initial learning rate& $10^{-3}$ \\ 
$\lambda$ &Weight decaying factor& $10^{-3}$ \\ 
$\vartheta$ &Base of the exponential factor& $10$\\
$\epsilon$ &Offset parameter of the exponential factor& $9$\\ 
\bottomrule
\end{tabular}
\label{tab:hyper}
\end{table}


The hyper-parameters settings for the training of DeepVLF are summarized in Table~\ref{tab:hyper}. To dynamically adjust its average coderate, we vary the decoding threshold $\gamma$ from $1-10^{-3}$ to $1-10^{-7}$. The parameter $\mu$ in \eqref{eq:tau+} is set to
\begin{equation}\label{eq:mu}
    \mu =
    \begin{cases}
     5,& \text{if $\gamma<=1-10^{-5}$;}\\
     6,& \text{if $1-10^{-5}<\gamma<=1-10^{-6}$;}\\
     7,& \text{if $\gamma>10^{-6}$.}
    \end{cases}
\end{equation}


\subsection{Performance}
Considering a short block length $K=51$ and noiseless feedback, we present the BLER versus code rate performances of DeepVLF in Figs.~\ref{fig:bler}(a) and~\ref{fig:bler}(b), where the forward SNRs are set to $1$ and $0$ dB, respectively. As can be seen, DeepVLF outperforms existing DL-aided feedback coding schemes, establishing a new state-of-the-art performance in high code rate scenarios. Specifically, under a code rate of $1/2$ and a forward SNR of $1$ dB, DeepVLF achieves a BLER below $10^{-5}$, whereas the best existing DL-based code, GBAF, reaches only up to a BLER of $10^{-2}$. This demonstrates DeepVLF's superior reliability and efficiency.

\begin{rem}
At such a short block length, conventional forward channel codes, such as Polar, LDPC, and Turbo codes, perform poorly compared to codes that leverage feedback.
\end{rem}


\begin{rem}
In LightCode \cite{ankireddy2024lightcode}, the definition of block error rate corresponds to the error rate of bit groups in this paper. Therefore, the BLER of LightCode shown in Fig.~\ref{fig:bler} is calculated based on the bit group error rates reported in \cite{ankireddy2024lightcode}.
\end{rem}

To further evaluate DeepVLF's performance, we extend our analysis to scenarios with noisy feedback. DeepVLF relies on two types of feedback: the received symbols and the indices of successfully decoded bit groups. As the indices comprise a minimal amount of data, we assume they are transmitted reliably over a dedicated control channel. Only the received symbol packets are sent over the noisy feedback channel. The results are presented in Fig.~\ref{fig:bler}(c), where the forward SNR is set to $1$ dB, and the feedback SNR is set to $20$ dB, consistent with prior studies. The experimental findings reveal that even under noisy feedback conditions, DeepVLF consistently outperforms other DL-based feedback coding schemes, especially in the high code rate regime. This demonstrates the robustness of DeepVLF to noise in the feedback channel, affirming its effectiveness in practical scenarios with feedback imperfections.

\subsection{Ablation study}
This section presents an ablation study to evaluate the contributions of VDFE and the loss function of DeepVLF, focusing on the $1$ dB forward channel SNR and noiseless feedback conditions.

\subsubsection{Impact of VDFE}
\begin{figure}[t]
    \centering
    \includegraphics[width=0.8\linewidth]{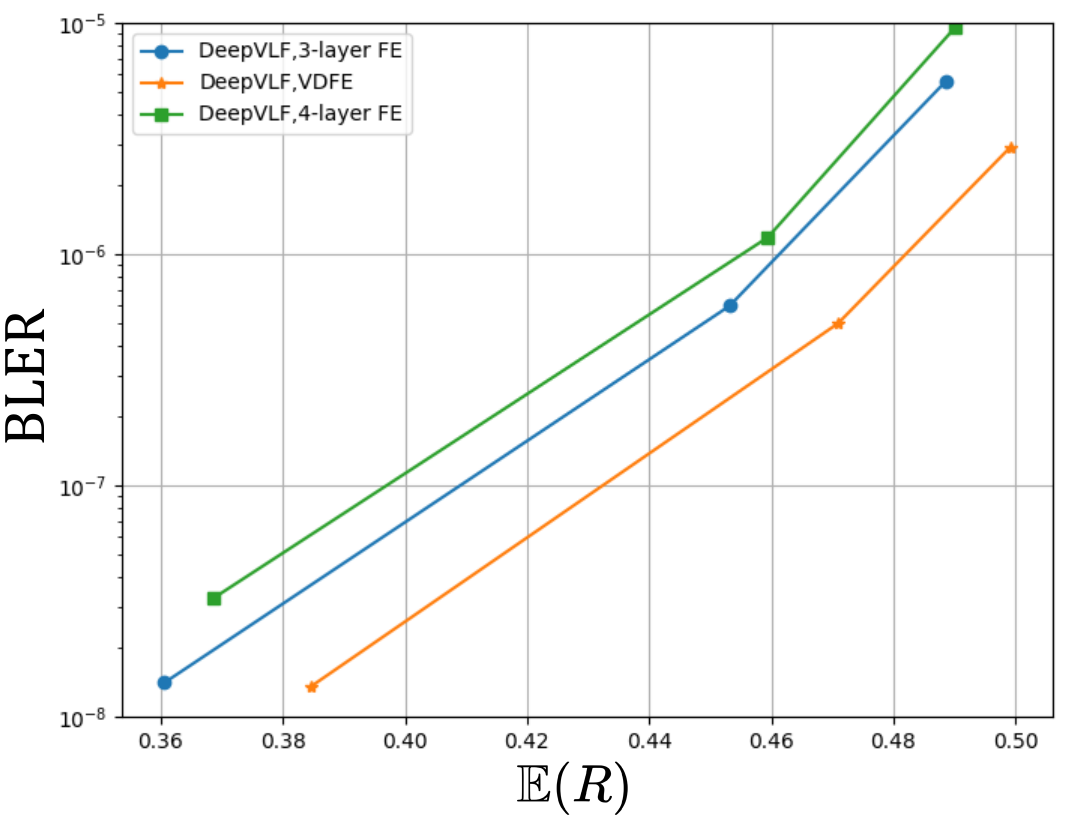}
    \caption{An ablation study to verify the effectiveness of VDFE.}
    \label{fe}
\end{figure}

In DeepVLF, the VDFE mechanism is implemented as a three-layer fully-connected network for $\tau \leq \tau_{\text{VD}}=3$. When $\tau>\tau_{\text{VD}}$, an additional linear layer with a ReLU activation function is incorporated, expanding VDFE to four layers. To validate the benefits of this variable-depth structure, we test fixed versions of the feature extractor with either three or four layers and compare their performances against the variable-depth configuration in DeepVLF. Fig.~\ref{fe} shows that the VDFE design in DeepVLF outperforms the fixed configurations, demonstrating its effectiveness.

\begin{table}[t]
\caption{The impact of loss function design.}
\centering
\setlength{\tabcolsep}{5mm} 
\begin{tabular}{ccc}
\toprule
\textbf{Loss Type} &\textbf{Code rate} & \textbf{BLER} \\ 
\midrule
Single  & 0.476 & $3.4\times10^{-6}$ \\ 
EqW & 0.481 &$9.6\times10^{-6}$ \\ 
DWA & 0.495 & $3.1\times10^{-6}$ \\ 
ExpW (ours) & 0.499 & $2.9\times10^{-6}$ \\
\bottomrule
\end{tabular}
\label{tab:loss}
\end{table}

\subsubsection{Impact of the loss function}
We examine the effectiveness of our customized loss function in \eqref{eq:loss2} by comparing it against several benchmark strategies:
\begin{itemize}
\item Single loss: use only the cross-entropy loss at the communication round where each bit group is decoded, i.e., $\tau^*_q$, as the loss function.
\item Equal weighting (EqW) loss: assign uniform weights to the losses across all communication rounds.
\item Dynamic weight averaging (DWA) loss \cite{bian2023deepjscc}: adapt weights dynamically based on the gradient of the classification header, allowing the model to focus on rounds with higher learning impact.
\item Exponential weighting (ExpW) loss: apply weights that  increase exponentially with the communication round $\tau$, giving greater emphasis to later rounds (i.e., our loss used in DeepVLF).
\end{itemize}

To compare the effectiveness of these schemes, we set $\gamma = 1-10^{-5}$ and evaluate their performance in terms of average code rate and BLER, as shown in Table \ref{tab:loss}. The results indicate that our loss function design achieves both the lowest BLER and the highest average code rate, demonstrating its superiority.

\section{Conclusion}
This work introduced DeepVLF, a DL-aided approach to variable-length feedback coding that dynamically adjusts code rates at the bit-group level by leveraging multi-round feedback. Through key innovations such as threshold-based decoding and adaptable parity symbol generation, DeepVLF offers significant improvements in decoding accuracy and resilience, even under high code rate requirements. Experimental results highlight DeepVLF's ability to outperform existing DL-aided feedback codes, establishing new benchmarks and demonstrating resilience to feedback noise.
Our work opens promising avenues for the future of adaptive communication systems. By further exploring the integration of DeepVLF into diverse network settings, particularly those with limited resources or variable channel conditions, the potential impact of this scheme can be expanded, paving the way for next-generation, ultra-reliable communication networks.


\bibliographystyle{IEEEtran}
\bibliography{References}

\end{document}